\documentclass[twocolumn]{aastex63}

\usepackage{color}

\shortauthors{Hu et al.}
\begin{document}

\title{A Channel to Form Fast-spinning Black Hole--Neutron Star Binary Mergers as Multi-messenger Sources}

\correspondingauthor{Ying Qin, Jin-Ping Zhu}
\email{yingqin2013@hotmail.com, zhujp@pku.edu.cn}

\author[0000-0002-6442-7850]{Rui-Chong Hu}
\affiliation{Guangxi Key Laboratory for Relativistic Astrophysics, School of Physical Science and Technology, Guangxi University, Nanning 530004, China}

\author[0000-0002-9195-4904]{Jin-Ping Zhu}
\affiliation{Department of Astronomy, School of Physics, Peking University, Beijing 100871, China}

\author[0000-0002-2956-8367]{Ying Qin}
\affiliation{Department of Physics, Anhui Normal University, Wuhu, Anhui, 241000, China}
\affiliation{Guangxi Key Laboratory for Relativistic Astrophysics, School of Physical Science and Technology, Guangxi University, Nanning 530004, China}

\author[0000-0002-9725-2524]{Bing Zhang}
\affiliation{Nevada Center for Astrophysics, University of Nevada, Las Vegas, NV 89154, USA}
\affiliation{Department of Physics and Astronomy, University of Nevada, Las Vegas, NV 89154, USA}

\author[0000-0002-7044-733X]{En-Wei Liang}
\affiliation{Guangxi Key Laboratory for Relativistic Astrophysics, School of Physical Science and Technology, Guangxi University, Nanning 530004, China}

 \author[0000-0003-2506-6906]{Yong Shao}
 \affiliation{Department of Astronomy, Nanjing University, Nanjing 210023, China}

\begin{abstract}
After the successful detection of a gravitational-wave (GW) signal and its associated electromagnetic (EM) counterparts from GW170817, neutron star--black hole (NSBH) mergers have been highly expected to be the next type of multi-messenger source. However, despite the detection of several of NSBH merger candidates during the GW third observation run, no confirmed EM counterparts from these sources have been identified. The most plausible explanation is that these NSBH merger candidates were plunging events mainly because the primary BHs had near-zero projected aligned-spins based on GW observations. In view that NSs can be easily tidally disrupted by BHs with high projected aligned-spins, we study an evolution channel to form NSBH binaries with fast-spinning BHs, the properties of BH mass and spin, and their associated tidal disruption probability. We find that if the NSs are born firstly, the companion helium stars would be tidally spun up efficiently, and would thus finally form fast-spinning BHs. If BHs do not receive significant natal kicks at birth, these NSBH binaries that can merge within the Hubble time would have BHs with the projected aligned-spins $\chi_{z}\gtrsim0.8$ and, hence, can certainly allow tidal disruption to happen. Even if significant BH kicks are considered for a small fraction of NSBH binaries, the projected aligned-spins of BHs are $\chi_z\gtrsim0.2$. These systems can still be disrupted events unless the NSs are very massive. Thus, NS-first-born NSBH mergers would be promising multi-messenger sources. We discuss various potential EM counterparts associated with these systems and their detectability in the upcoming fourth observation run.

\end{abstract}


\section{Introduction}

Neutron star--black hole (NSBH) mergers are prime search targets for the ground-based gravitational wave (GW) detectors, e.g., LIGO \citep{aasi2015}, Virgo \citep{acernese2015} and KAGRA \citep{aso2013}. Recently, two high-confidence GW events (GW200105 and GW200115) were announced by the LIGO-Virgo-KAGRA (LVK) Collaboration, which was, for the first time, identified to come from mergers of NSBH binaries \citep{abbott2021observation,nitz2021}. Furthermore, two lower mass-gap sources (GW190814 and GW200210\_092254) that could either be from a NSBH or a binary BH, and several marginal NSBH candidates were discovered in GW during the third observation run (O3) of LVK \citep{abbott2020GW190814,abbott2021gwtc2,abbott2021gwtc21,abbott2021gwtc3}. The analysis of these NSBH candidates displayed that their effective inspiral spins could be near-zero or very low, suggesting that their primary BHs would plausibly have negligible projected spins aligned to the direction of the orbital angular momentum (abbreviated as projected aligned-spins, hereafter). Among these candidates, only the BH component of GW200115 showed a non-zero BH spin with an apparent BH spin-orbit misalignment angle \citep{abbott2021observation}, which may require a strong natal kick for the BH or the NS \citep{fragione2021impact,gompertz2021,zhuxj2021}. Conversely, \cite{mandel2021GW200115} argued that GW200115 could potentially be a merger between a non-spinning BH and a typical mass of NS by reanalysing its GW signal with the consideration of astrophysically motivated priors.

NSBH mergers have long been proposed to be progenitors of some fast-evolving electromagnetic (EM) counterparts, including short-duration gamma-ray bursts \citep[sGRBs;][]{paczynski1991,narayan1992,zhang2018} and kilonovae \citep{li1998,metzger2010}. As main GW sources of ground-based GW detectors, it was thus expected that the follow-up searches after GW triggers could help find out the associated EM signals of NSBH mergers. However, despite many efforts toward the follow-up observations, no confirmed EM counterpart candidate was identified \citep[e.g.,][]{anand2021,gompertz2020,kasliwal2020,coughlin2020implications1,coughlin2020implication2,page2020,becerra2021}. The tidal disruption probability of NSBH mergers and the brightness of EM signals depend on NS mass, NS equation of state (EoS), BH mass, and especially BH projected aligned-spin \citep[e.g.,][]{kyutoku2015,kawaguchi2016,foucart2018,barbieri2019,raaijmakers2021,zhu2020,zhu2021population,zhu2021no,zhu2021kilonova}. Disrupted events associated with brighter EM signals tend to occur only if a NSBH binary has a low-mass NS component with a stiff EoS and a low-mass BH component with a high projected aligned-spin. The most promising explanation for the lack of EM counterparts is that these NSBH candidates were plunging events without forming any bright EM signals, mainly due to their near-zero BH projected aligned-spins \citep{zhu2021no,fragione2021NSBH,dorazio2021}.

The most widely accepted formation channel for the majority of NSBH binaries in the universe is the classical Common Envelope (CE) scenario \citep[e.g.,][]{giacobbo2018,Belczynski2020,drozda2020,Shao2021,broekgaarden2021impart}. In this scenario, the immediate progenitor of a NSBH binary just after the CE phase is a close binary system, which consists of a compact object (NS or BH) and a helium star. On the one hand, if the first-born compact object is a BH, its spin has been found to be negligible \citep{Qin2018,Belczynski2020}. This was supported by the finding of the near-zero distribution of BH spin in the O3 GW NSBH candidates \citep{zhu2021population}, which implies that the NSs in these NSBH candidate systems may be directly plunged into the BHs. On the other hand, for a fraction of NSBH progenitor systems in which the NS was first born, the companion helium star can be spun up by the NS and finally form a BH with a high projected aligned-spin. Since a fast-spinning BH can easily tidally disrupt the NS and produce bright EM signals (i.e., sGRBs and kilonovae), NSBH mergers formed via this  formation channel can be potential  multi-messenger sources that allow us to discover their associated bright EM counterparts with a high probability after GW triggers \citep{zhu2021kilonova}. Furthermore, as studied recently by \citet{Jaime2021},  core-collapse physics plays a critical role in the observability of the EM signals produced by NSBH mergers. In viewing of the lack of relevant researches on the BH spin properties of such NS-first-born formation channel for NSBH mergers, we investigate this formation channel of NSBHs with fast-spinning BHs in detail and study their corresponding tidal disruption probability.

In this work, we investigate the detailed binary evolution process of forming fast-spinning BHs in NSBH binaries by taking into account the accretion feedback of the core-collapse processes, supernova (SN) kicks of newly-formed BHs, and different NS EoSs. The main methods adopted in the stellar and binary evolution models are shown in Section \ref{sec:2}. We then present in Section \ref{sec:3} our findings of NSBH mergers without and with natal kicks, respectively, along with their associated probability for tidal disruption. The multi-messenger observational signatures of this channel are discussed in Section \ref{sec:4}. Finally, main conclusions are summarized in Section \ref{sec:5} with some discussion.

\section{Methods} \label{sec:2}
We use the release 15140 of \texttt{MESA} stellar evolution code \citep{Paxton2011,Paxton2013,Paxton2015,Paxton2018,Paxton2019} to perform all of the binary evolution calculations in this work. A metallicity of $Z = Z_{\odot}$, where the solar metallicity is $Z_{\odot} = 0.0142$ \citep{Asplund2009}, is adopted. We create single helium stars at Zero-Age Helium Main-Sequence (ZAHeMS) following the method in \cite{Qin2018} and then relax the created helium stars to reach the thermal equilibrium, where the He-burning luminosity just exceeds 99\% of the total luminosity. We model convection using the mixing-length theory \citep{MLT1958} with a mixing-length parameter $\alpha = 1.93$. The Ledoux criterion is used to treat the boundaries of the convective zone, while we consider the step overshooting as an extension given by $\alpha_p = 0.1 H_p$, where $H_p$ is the pressure scale height at the Ledoux boundary limit. Semiconvection \citep{Langer1983} with an efficiency parameter $\alpha_{\sc} = 1.0$ is also included in our model. The network of \texttt{approx12.net} is chosen for nucleosynthesis.

Stellar winds are modelled with the standard ``\texttt{Dutch}'' scheme, calibrated by multiplying with a scaling factor of 0.667 to match the recently updated modeling of helium stars' winds \citep{Higgins2021}. We model angular momentum transport and rotational mixing diffusive processes \citep{Heger2000}, including the effects of Eddington–Sweet circulations, the Goldreich–Schubert–Fricke instability, as well as secular and dynamical shear mixing. We adopt diffusive element mixing from these processes with an efficiency parameter of $f_c = 1/30$ \citep{Chaboyer1992,Heger2000}. Mass transfer is modeled following the Kolb scheme \citep{Kolb1990} and the implicit mass transfer method \citep{Paxton2015} is adopted.

We model helium stars until the carbon depletion in the center. The baryonic remnant mass is calculated following the ``delayed'' supernova prescription in \citet{Fryer2012}. As shown in \citet{Batta2019}, the newly formed BH might be unable to accrete all of the available stellar material. Therefore, in order to calculate the final mass and spin of the BH from the direct collapse, we follow the framework given in \citet{Batta2019}, which has already been implemented in a recent work \citep{Bavera2020}. The neutrino loss as in \citet{Zevin2020} is taken into account. Furthermore, the maximum NS mass is assumed to be $2.5\,M_{\odot}$ in this work.

BHs formed through direct collapse are considered to receive no mass loss and thus no natal kick \citep{Belczynski2008}. Recently, an estimation of the NSBH merger rate from the LIGO-Virgo O3a run showed that negligible kicks imparted on the low-mass BHs is favored \citep{Jaime2021}. Consequently, calculation with no natal kick included is considered as our fiducial model. Additionally, we also take into account natal kicks onto BHs formed through direct collapse. We follow the parametrised recipes of \cite{Mandel2020recipe} to calculate the natal kicks of BHs. The binary properties just after the natal kick is given based on the framework of \cite{Vicky1996,Wong2012} and a recent work by \citet{Callister2021}. For binaries surviving after the kick, we estimate the merger time given from \citet{Peters1964} through GW emission, with the updated fitting formula for eccentricity in \cite{Mandel2021ecc}.

\section{Results} \label{sec:3}

Here we present a parameter space study of the initial binary properties for a close binary system composed of a helium star and a NS. The parameters include the initial helium star masses and the initial orbital periods. We cover the helium star mass range $8-40\,M_\odot$, the NS star mass range $1.2-2.5\,M_\odot$, as well as the orbital period from 0.2 to 2\,days. We also take into account the natal kicks imparted onto BHs in the core-collapse models. We describe in detail our findings for the NSBH formation without and with natal kicks, respectively, along with their associated tidal disruption probability. 

\subsection{Fast-spinning BHs originated from tidal spin-up in NSBH binaries}

 \begin{figure*}[htbp]
      \centering
      \includegraphics[width=0.995\columnwidth]{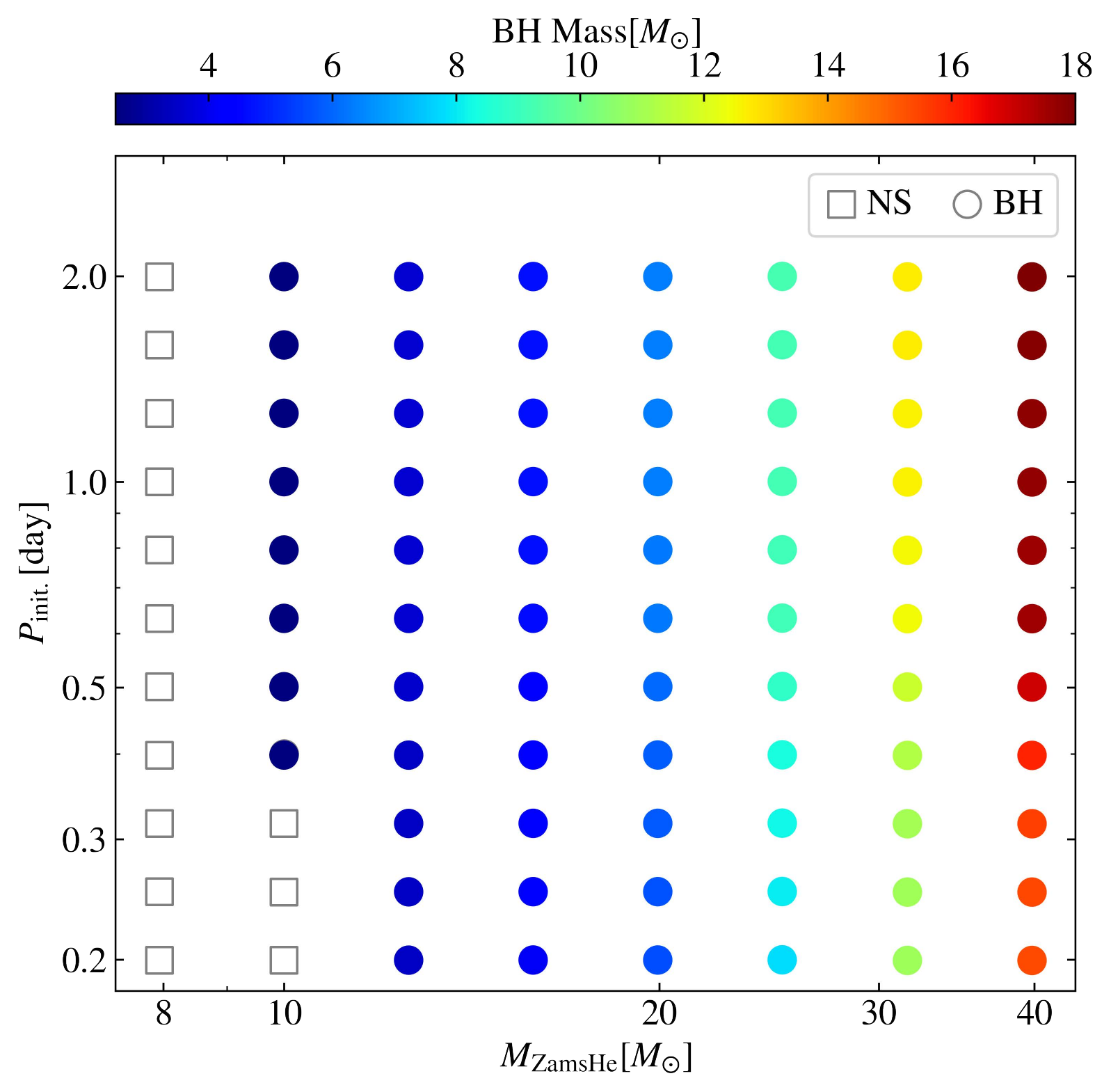}
      \includegraphics[width=\columnwidth]{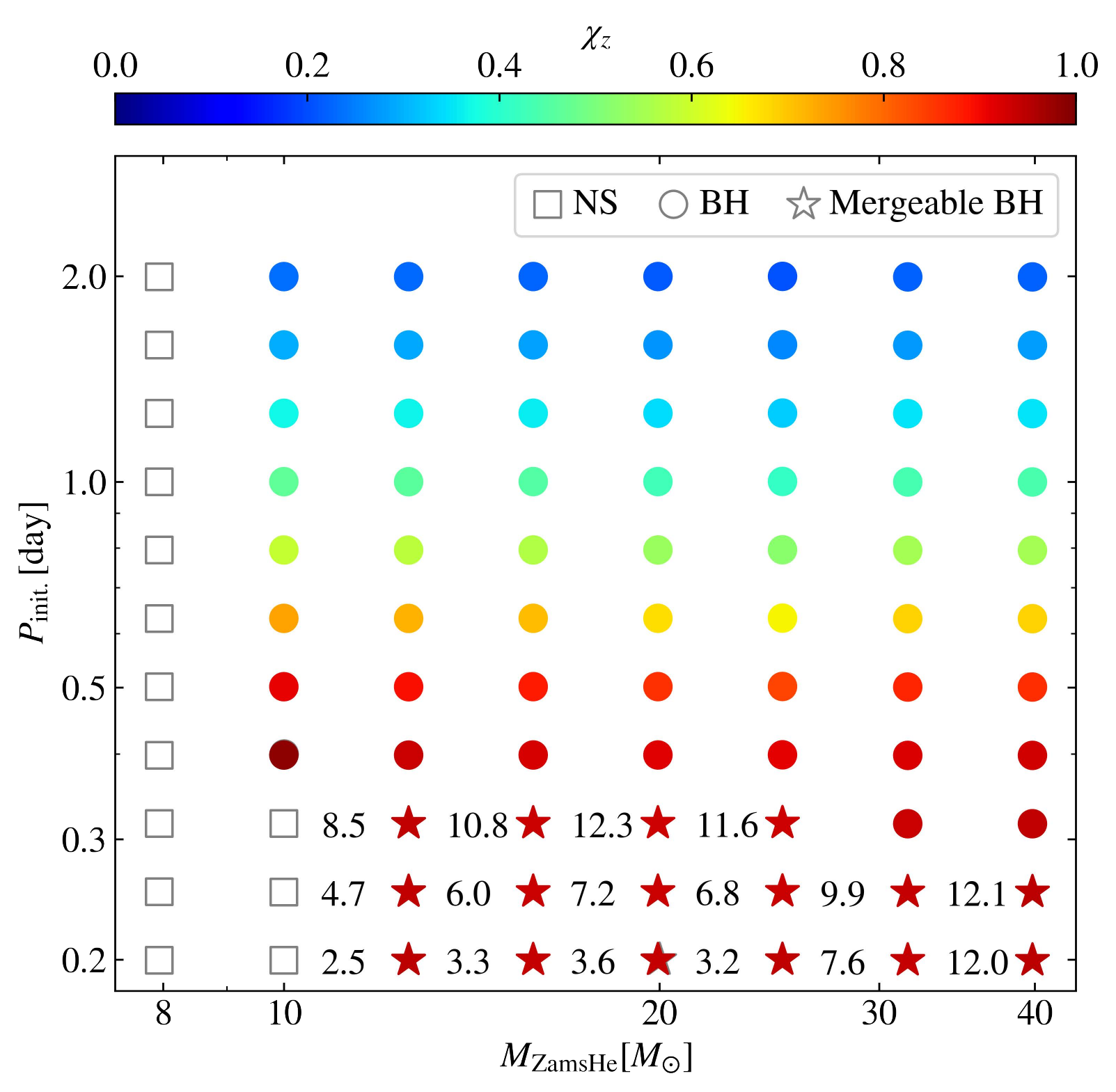}
      \caption{Left panel: BH mass (the color bar) as a function of the helium star initial mass and orbital period. The symbols represent different compact objects formed through direct core-collapse of the helium stars, i.e., squares for NS and dots for BH. Right panel: same as the left panel, but the color represents the BH projected aligned-spins. The star symbols are marked with NSBH binaries that can be merged via GW emission within Hubble time. The numbers on the left side of the star symbols are the corresponding merger time in units of billion years.}
      \label{a1}
 \end{figure*}

Figure \ref{a1} presents various outcomes of the detailed binary evolution for helium stars at different initial masses orbiting around a NS of $1.4\,M_\odot$ in close orbits. We choose 2\,days as the upper limit of the orbital period, beyond which tides are not important \citep{Qin2018}. We do not take into account the parameter space where the initial orbital periods are shorter than 0.2\,days, as the binary would experience either initial overflow or dynamically unstable mass transfer. NSs with other masses have similar results, which are not shown in the paper.

First, we note from the left panel of Figure \ref{a1} that helium stars with the initial mass higher than $10\,M_\odot$ and initial orbital period larger than $\sim0.4$\,days collapse to form BHs. Furthermore, helium stars in tighter orbits tend to lose more mass due to rotationally enhanced mass loss \citep{Heger1998,Langer1998}. This is why we can see that helium stars in short orbits end up forming low-mass BHs. For low-mass ($<10\,M_\odot$) helium stars, including a small fraction with initial orbital periods less than $\sim0.3$\,days, NSs are formed instead. A further investigation of binary NS systems formed in our grid is postponed in a future work.

In the right panel of Figure \ref{a1}, we show the magnitudes of the BH projected aligned-spin for different initial binary properties. As shown in \cite{Qin2018}, the spins of the resulting second-born BHs are exclusively dependent on the tidal interaction and stellar winds of the helium stars. For initial orbital periods considered in this gird, the BHs have the entire range of the spin from zero to value allowed by maximally spinning. For more massive helium stars, their stellar winds are stronger, which widen the binaries and thus make NSBHs undetectable for GW emission within Hubble Time. Interestingly, we note that for helium stars with different initial masses, the BH projected aligned-spins have a similar trend in decreasing magnitudes. After the formation of the second-born BH, the GW emission removes the orbital angular momentum of the NSBH, shrinks the orbit and eventually leads to merger of the two compact objects. The merger time is calculated as given in \cite{Peters1964}. It is worth noting that all NSBHs with their merger time less than the Hubble time have extremely fast spins. This finding is mainly attributed to the strong tidal force in close binaries. However, BHs formed through direct core-collapse may receive SN kicks, which can change the direction of the BH spins and, hence, their projected aligned-spins. The impact of SN kicks is discussed as follows.

\subsection{Impact of SN kicks on BH projected aligned-spins and the probability of merging NSBHs} \label{sec:kick}

 \begin{figure*}[htbp]
      \centering
      \includegraphics[width=\columnwidth]{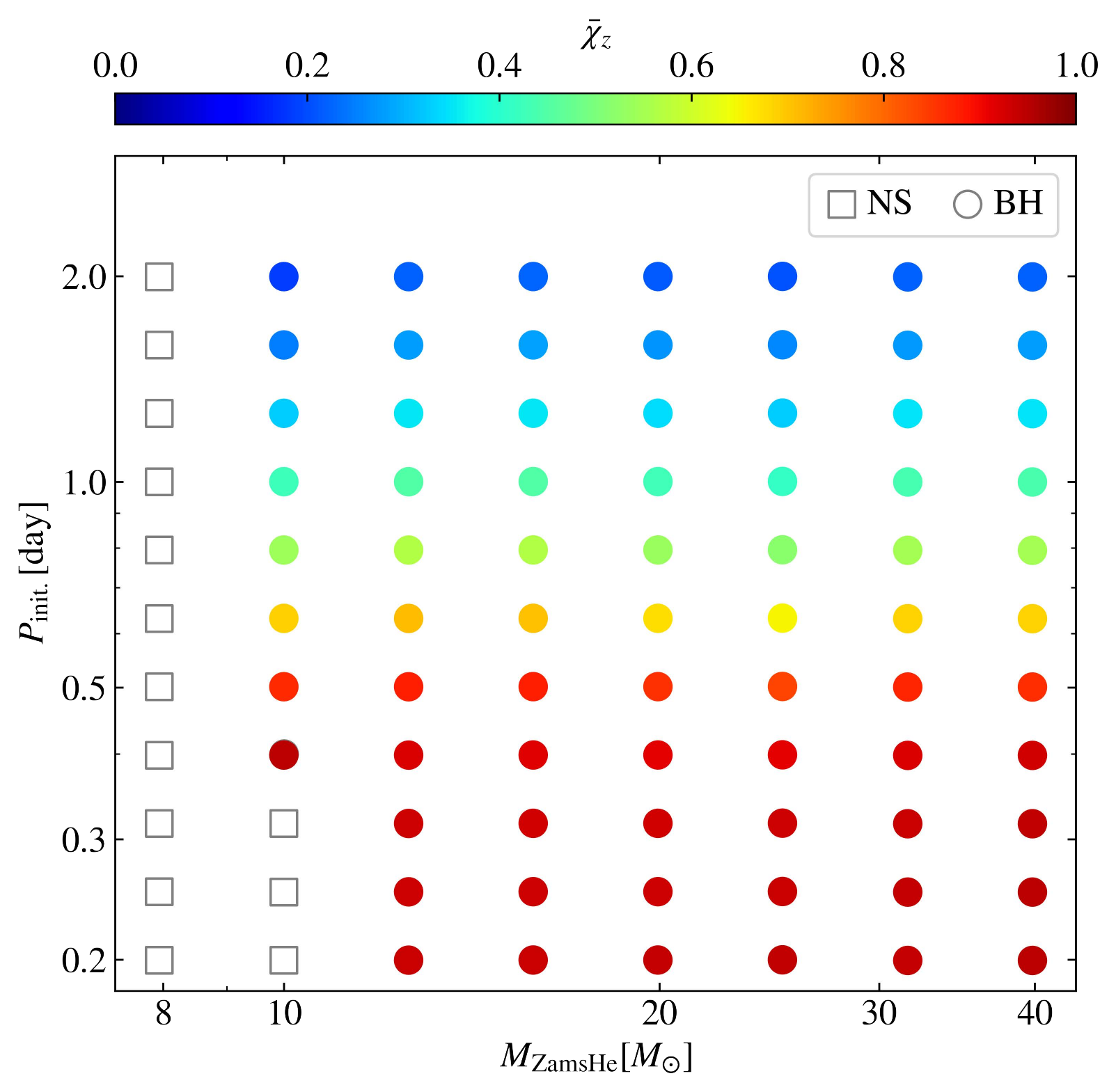}
      \includegraphics[width=\columnwidth]{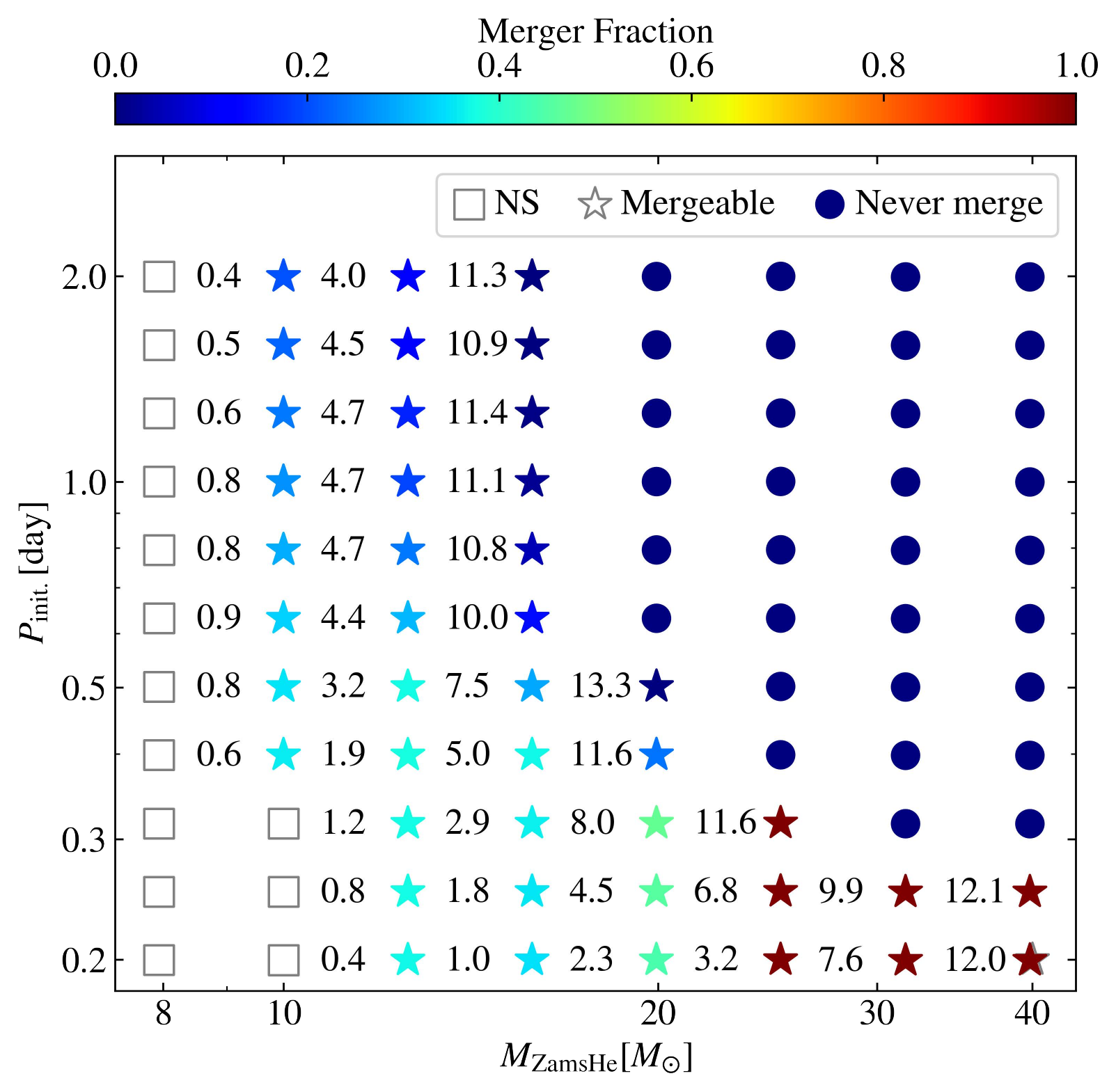}
      \caption{Left panel: similar to the left panel in Figure \ref{a1}, but with SN kicks included. Right panel: the color represents the fraction of NSBH binaries which can be merged within Hubble time with SN kicks considered. The numbers on the left side of the star symbols are the medians of the corresponding merger time in units of billion years that is less than Hubble time. Dark blue dots: NSBH binaries that never merge within Hubble time.}
      \label{a2}
 \end{figure*}

The SN kicks imparted onto BHs are considered to cause a misalignment of the BH spin to the orbital angular momentum. In addition, the post-SN orbital separation and the eccentricity of the binary can also be modified. We briefly describe these calculations in Appendix and more details can be found in \cite{Vicky1996,Wong2012,Callister2021}. Similar to \cite{Mandel2020recipe}, kicks are drawn from a Gaussian distribution and we repeat $10^5$ times to obtain the average value of the kick and also its associated angle $\theta$. The corresponding averaged BH projected aligned-spins ($\bar{\chi}_{z}$) are shown in the left panel of Figure \ref{a2}. We note that $\bar{\chi}_{z}$ values are slightly less than those obtained assuming no kicks (see the right panel of Figure \ref{a1}). This is because the SN kicks we apply to the BHs are typically small \citep[see the red line for BH kicks in Figure 4 of][]{mandel2021GW200115}. Accordingly, we then obtain on average a small misalignment for BH spins.

Let us see how kicks change the fraction of merging NSBHs. As shown in the right panel of Figure \ref{a2}, with the post-SN-explosion binary properties updated, for each initial system we can then calculate the fraction of the post-SN-explosion NSBH systems that can survive and merge within Hubble time. The BH masses of these systems against their BH projected aligned-spins are presented in Figure \ref{fig:TidalDisruption}. As the kick is considered, massive BHs with $M_{\rm BH}\gtrsim 7.6\,M_\odot$ do not experience natal kicks (the last three columns from the right side) as they are heavier than the carbon-oxygen core of their progenitors at the pre-SN state. Some NSBH binaries with relatively low spins, which can not merge in a wide orbit, can instead merge within Hubble time caused by the SN kicks. However, the merger fraction deceases since the systems have lower-mass BHs with lower projected aligned-spins. More specifically, as shown in Figure \ref{fig:TidalDisruption}, BHs with $M_{\rm BH}\lesssim 5\,M_\odot$ have a wide range of the projected aligned-spin $\chi_z$ from $\sim0.2$ to $\sim1.0$, while more massive BHs are found to be extremely spinning, i.e., $\chi_z\gtrsim0.8$.

The first-born BHs in NSBH mergers are expected to have near-zero spin distributions \citep[e.g.,][]{Qin2018,Belczynski2020}. \cite{broekgaarden2021formation,kinugawa2022} suggested that GW200105 and GW200115 are expected to be formed via the classical BH-first-born isolated formation channel. Their posterior distributions of BH projected aligned-spin versus BH mass from \cite{abbott2021observation} are plotted in Figure \ref{fig:TidalDisruption}. It is clear that the distributions of the BH projected aligned-spin for GW200105 and GW200115 are quite different from our calculated distributions for NS-first-born NSBH mergers.

\subsection{Tidal disruption probability} \label{sec:3.3}

\begin{figure*}[htpb]
    \centering
    \includegraphics[width = 0.49\linewidth , trim = 40 35 130 30, clip]{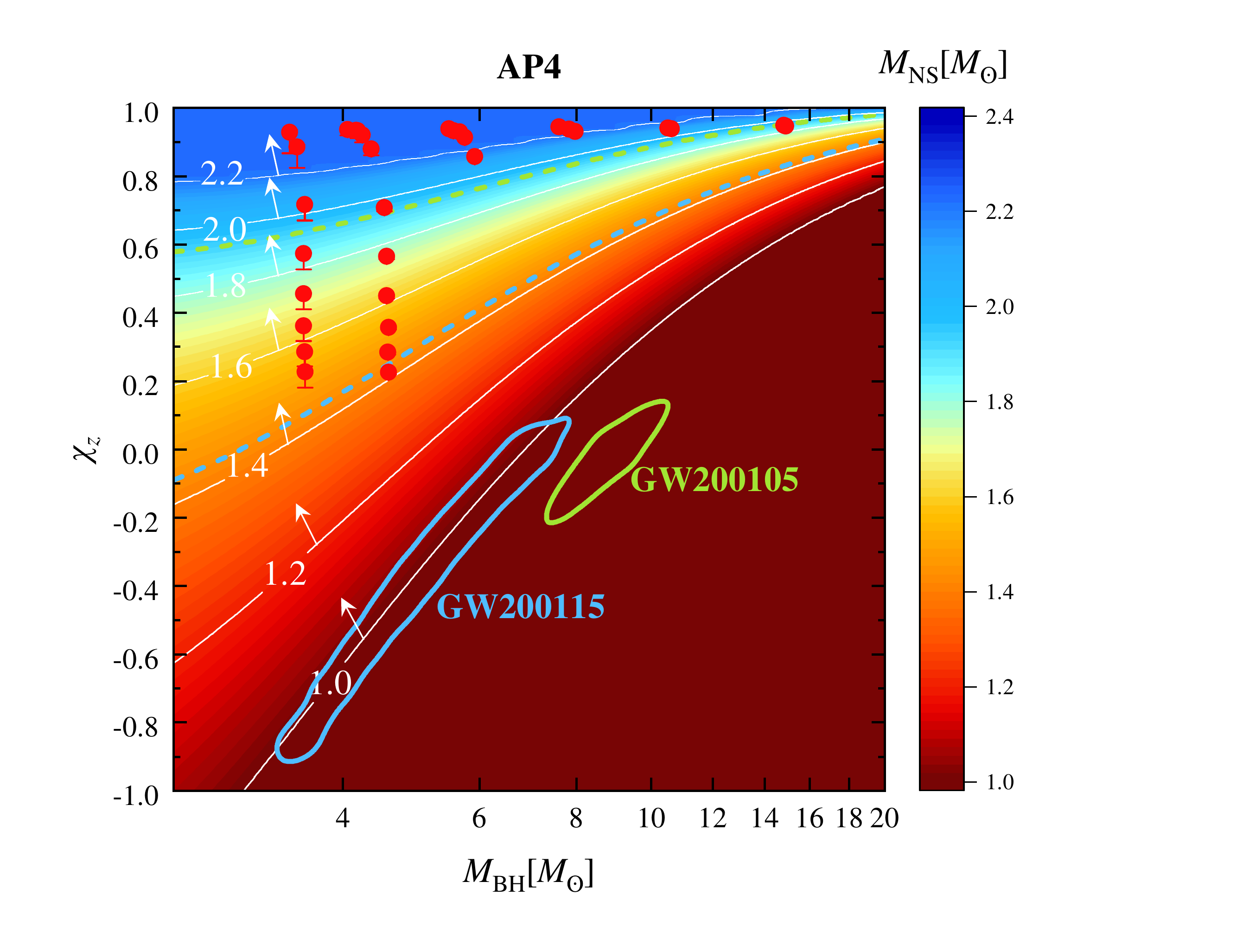}
    \includegraphics[width = 0.49\linewidth , trim = 40 35 130 30, clip]{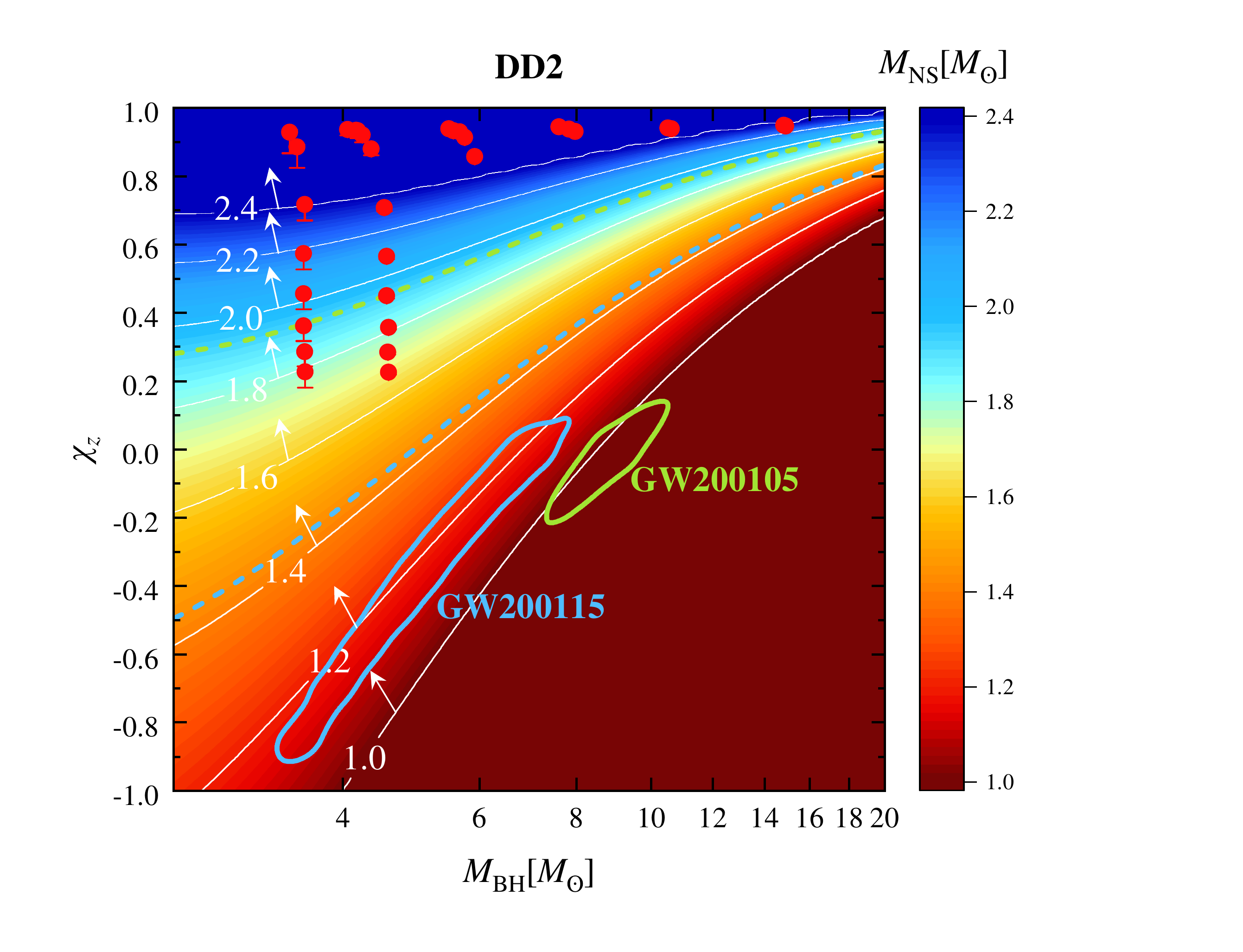}
    \caption{The median and $90\%$ confidence interval of BH projected aligned-spin versus BH mass from the grid of calculations for those merger events within Hubble time used to produce Figure \ref{a2} (red circles), and the parameter space for NSBH merger systems to allow tidal disruption of the NS by the BH. Two EoSs, i.e., AP4 (left panel) and DD2 (right panel), are considered. We mark several values of the NS mass from $M_{\rm NS} = 1\,M_\odot$ to $M_{\rm NS} = 2.2\,M_\odot$ for AP4 ($M_{\rm NS} = 2.4\,M_\odot$ for DD2) as solid lines in each panel. For a specific $M_{\rm NS}$, the NSBH mergers with BH mass and BH projected aligned-spin located at the top left parameter space (denoted by the direction of the arrows) can allow tidal disruptions to occur. For GW200105 (green) and GW200115 (blue), the $90\%$ credible posterior distributions (colored solid lines) of the parameters obtained from \cite{abbott2021observation} are displayed, respectively. Corresponding median values of $M_{\rm NS}$ for these two sources are marked as dashed lines. Both GW200105 and GW200115, which are believed to form via the BH-first-born formation scenario \citep{broekgaarden2021formation,kinugawa2022}, are most likely plunging events.}
    \label{fig:TidalDisruption}
\end{figure*}

The tidal disruption probability of NSBH mergers is determined by NS mass, NS EoS, BH mass, and BH projected aligned-spin. By considering two representative EoSs, i.e., AP4 \citep{akmal1997} and DD2 \citep{typel2010}, in which AP4 is one of the most likely EoSs while DD2 is one of the stiffest EoSs constrained by GW170817 \citep{abbott2018GW170817,abbott2019properties}, the parameter space where the NS can be tidally disrupted using the formula from \cite{foucart2018} is shown in Figure \ref{fig:TidalDisruption}. A NSBH binary system that has a lower-mass NS with a stiffer EoS and a lower-mass BH with a higher projected aligned-spin can more easily make tidal disruption.
The Tolman-Oppenheimer-Volkov (TOV) mass for EoSs of AP4 and DD2 is $M_{\rm TOV} = 2.22\,M_\odot$ and $2.42\,M_\odot$, respectively. On one hand, for almost all of the parameter space of the NS mass, Figure \ref{fig:TidalDisruption} reveals that NSBH mergers formed via NS-first-born formation scenario should be disrupted events if the remnant BHs are heavier than $\sim5\,M_\odot$. Furthermore, for the BH mass located in the range of $M_{\rm BH} \lesssim 5\,M_\odot$, NSBH mergers whose BH projected aligned-spins are $\chi_{z}>0.8$ are expected to make tidal disruption as shown in Figure \ref{fig:TidalDisruption}. On the other hand, some mergers of NSBH binaries with relatively low BH projected aligned-spins (i.e., $\chi_{z}\lesssim0.8$) caused by the impact of SN kicks on low-mass BH components could still be plunging events, if the mass of the NS companions is $M_{\rm NS} \gtrsim 1.5\,M_\odot$ ($M_{\rm NS} \gtrsim 1.7\,M_\odot$) with the adoption of an EoS of AP4 (DD2). However, since the fraction of NSBH binaries with low BH projected aligned-spins that can merge within  Hubble time after the SN kick is limited as discussed in Section \ref{sec:kick}, plunging events may account for only a small part of NSBH mergers formed via NS-first-born formation scenario. Therefore, regardless of the EoS we choose, it is expected that most NSs can be tidally disrupted by the BHs in NS-first-born NSBH merger systems and they would be plausible  multi-messenger sources, which may be discovered in the future.

\section{Observational Signatures} \label{sec:4}

\subsection{Multi-messenger signals}

In Section \ref{sec:3.3}, we showed that most of NS-first-born NSBH mergers can easily make tidal disruption. Disrupted NSBH mergers are believed to drive some bright EM counterparts. After a NSBH merger, there might be a pair of relativistic jets launched along the polar axis by the accreting remnant BH via the Blandford--Znajek effect \citep{blandford1977,gompertz2020NSBH}, which can power a sGRB and its broad-band afterglow emission as the jet interacts with the interstellar medium. The radioactive decays of the rapid neutron-capture-formed heavy nuclei can effectively heat the ejected materials in the dynamically ejecta ejecta or the wind outflows launched from the disk around the remnant BH, which would power the fast-evolving kilonova emission \citep[e.g.,][]{kawaguchi2016,barbieri2019,zhu2020,darbha2021}. The brightness of the emission may be further enhanced by the energy injection from the central engine \citep[e.g.,][]{ma2018,qi2021}.

In the fourth observing run (O4), the number of detected NSBH GWs is believed to increase significantly \citep[e.g.,][]{abbott2020properties,zhu2021kilonova}. A higher accuracy in both localization and distance measurement by the GW observations is expected, which make the post-GW-trigger follow-up search observations more efficiently. Furthermore, during the same GW period, several more powerful telescopes, e.g., the Space Variable Objects Monitor \citep{wei2016} in gamma-rays, the Einstein Probe \citep{yuan2016} in X-rays, the Large Synoptic Survey Telescope \citep{lsst2009} and the Wide-Field Infrared Transient Explorer \citep{frostig2021} in optical-infrared bands, will start to operate and join the GW follow-up observational campaigns. It is possible that multi-messenger signals from NSBH mergers, especially for NS-first-born NSBH mergers, could be discovered in the forthcoming O4.

\subsection{A Possible Connection Between Long-duration and Short-duration GRBs}

One interesting inference from this channel is that there could be a common progenitor system for a small fraction of long-duration GRBs (lGRBs) and sGRBs. As shown in Section \ref{sec:3}, the resulting BHs in close binaries are found to be highly spinning due to tidally spin-up. Such a fast-spinning BH can in principle launch a relativistic jet and  produce a lGRB through the Blandford-Znajek mechanism \citep{blandford1977} during the formation of the second-born BH. After the formation of the BH, the NSBH binary would merge long time later after losing orbital energy and angular momentum due to GW emission. Eventually, a sGRB may be produced at the merger.

We show for each NSBH system the merger time, namely the time delay between the lGRB and the following sGRB, on the left side of the symbol in the right panel of Figure \ref{a1}. The time delay varies from a few tens to hundred billion years. When the SN kicks imparted onto the newly-formed BHs are considered, the average merger time\footnote{We adopt the median of the merger time in $10^5$ times drawings as the average merger time. We exclude the systems that are disrupted or whose merger times are beyond Hubble time.} drops by roughly one to two orders of magnitude (see the right panel of Figure \ref{a2}. In any case, the delay time is too long and it is impossible to directly test observationally the existence of both a lGRB and a sGRB from the same progenitor system. One possible approach to test this scenario is through host galaxy studies. It has been well established that lGRBs and sGRBs have statistically very different host properties \citep[e.g.][]{fruchter06,fong10,liy16}. Cross comparing the host properties with the consideration of galaxy evolution in the timescale of the delay between the SN explosion and merger may offer some clues. This is beyond the scope of this paper.

\section{Conclusions and Discussion} \label{sec:5}

In this work, we first present a detailed binary evolution modeling of the formation of NSBH systems, concentrating on a special channel in which NSs are first formed and BHs are born with high natal spins. With this formation scenario considered, we explore the tidal disruption probability of NSBH mergers, taking into account the impact of accretion feedback of direct core-collapse modeling on the mass and spin of the newly-formed BHs, SN kicks, as well as different NS EoSs. We find that these NSBH mergers produce BHs with extremely high natal spins. With no natal kicks for BHs, we note that NSBH binaries that can merge within Hubble time would have BHs with the projected aligned-spins $\chi_{z}\gtrsim0.8$, and therefore, can definitely have tidal disruption of the NSs at the merger. On the other hand, when natal kicks are taken into account, BHs with $M_{\rm BH}\lesssim 5\,M_\odot$ and with projected aligned-spins $\chi_z$ down to 0.2 (see the right panel of Figure \ref{a2}) can still merge with a NS within Hubble time, but with a low probability. These systems can still be disrupted events to produce bright EM signals unless the NSs are very massive.

Very recently, \cite{Olsen2022} reported an marginal candidate, GW190920\_113516, whose secondary could be a heavy NS. With an effective inspiral spin of $\chi_{\rm eff} = 0.60^{+0.26}_{-0.07}$, this event could be a potential NSBH merger with the NS first born. \cite{Jaime2021} claimed that the fraction of the NS-first-born systems with different SN engines is $\sim10\%$ \citep[see also][]{kinugawa2022}. The event rate and system parameter distribution for NS-first-born NSBH mergers are subject to further studies in the future. With the upgrade of GW observatories and the update of large survey telescopes, it is foreseen that one may detect high-confidence GW signals with associated EM counterparts from NSBH mergers, especially for NS-first-born NSBH mergers, in the near future GW-led multi-messenger era.

\acknowledgments
YQ acknowledges the support from the Doctoral research start-up funding of Anhui Normal University and from Key Laboratory for Relativistic Astrophysics in Guangxi University.
EWL is supported by the National Natural Science Foundation of China (Grant Nos. 12133003, U1731239) and the Guangxi Science Foundation (Grant No. AD17129006).
This work is supported by the National Natural Science Foundation of China (Grant Nos. 12192220, 12192221) and by the Natural Science Foundation of Universities in Anhui Province (Grant No. KJ2021A0106).

\clearpage
\appendix
Our implementation for natal kicks, imparted on the newly formed BH, is based on the framework of \citet{Vicky1996,Wong2012} and a recent work in \citet{Callister2021}. Following the prescription in \cite{Mandel2020recipe}, the mean natal kick for BHs is

\begin{equation}
    \mu_{\mathrm{kick}}=v_{\mathrm{BH}} \frac{\max \left(M_{\mathrm{CO}}-M_{\mathrm{BH}}, 0\right)}{M_{\mathrm{BH}}},
    \label{kick}
\end{equation}
where $v_{\mathrm{BH}}$ is the BH scaling prefactor and $M_{\mathrm{CO}}$ is the carbon-oxygen core (CO core) mass, respectively. We adopt $v_{\mathrm{BH}}=200\,\mathrm{km\,s^{-1}}$ as suggested in \cite{Mandel2020recipe}. We also add an additional component drawn from a Gaussian distribution with the standard deviation $0.3\mu_{\mathrm{kick}}$.

We update the semi-major axis and the eccentricity of the binary after the supernova (SN) kick following \cite{Vicky1996}
\begin{equation}
\begin{array}{r}
    a_{f}=G\left(M_{\rm NS}+M_{\rm BH}\right)\bigg[\displaystyle\frac{2 G\left(M_{\rm NS}+M_{\rm BH}\right)}{a_{i}}-v_{\text {kick }}^{2}-v_{\text {orb}}^{2} -2 v_{\text {kick},y} v_{\text {orb}}\bigg]^{-1}
\end{array}
\end{equation}
and
\begin{equation}
    1-e^{2}=\frac{\left(v_{\text {kick},y}^{2}+v_{\text  {kick},z}^{2}+v_{\text {orb}}^{2}+2 v_{\text {kick},y} v_{\text {orb}}\right) a_{i}^{2}}{G\left(M_{\rm NS}+M_{\rm BH}\right) a_{f}}
\end{equation}
where $G$ is Newton's constant, $a_{i}$ and $a_{f}$ are the pre- and post-SN semi-major axis, and $v_{\mathrm{kick},i}$ represents the $i$-th component of the kick velocity $\mathbf{v}_{\text {kick}}$. Here $v_{\text {orb}}$, the pre-SN orbital velocity of the BH progenitor relative to its companion (NS), is given \citep{Wong2012} by

\begin{equation}
    v_{\text {orb}}^2=G\left(M_{\rm NS}+M_{\rm pre-SN}\right)\left(\frac{2}{r}-\frac{1}{a_{i}}\right) ,
\end{equation}
where $r$ is the orbital separation between the BH progenitor and its companion at the pre-SN stage and $M_{\rm pre-SN}$ ($M_{\rm He}$ at the pre-SN phase) is the mass of the BH progenitor just prior to the SN explosion. The expression of $r$ is given in Equation (15) in \cite{Wong2012}.

Following the same assumption \citep{Vicky1996} that the explosion is instantaneous, $r$ remains unchanged. As a result, the angle $\theta$ between the pre- and post-SN orbital planes is determined by

\begin{equation}
    \cos \theta=\frac{v_{\text{kick}, y}+v_{\text {orb}}}{\left[\left(v_{\text {kick}, y}+v_{\text {orb}}\right)^{2}+v_{\text {kick}, z}^{2}\right]^{1 / 2}} .
\end{equation}

If the natal kick is too strong, the binary after the supernova explosion will be disrupted. As shown in \citet{Callister2021}, the binary is disrupted when

\begin{equation}
    \beta<\frac{1}{2}+\frac{v_{\text {kick }}^{2}}{2 v_{\text {orb }}^{2}}+\frac{\mathbf{v}_{\text {kick }} \cdot \mathbf{v}_{\text {orb }}}{v_{\text {orb }}^{2}} ,
\end{equation}
where $\beta$ is defined in Equation (\ref{equ:7}) in \cite{Vicky1996}, the pre-SN-to-post-SN ratio of the binary's total mass. For binaries surviving after the SN kicks, we can calculate the merger time via gravitational-wave emission \citep{Mandel2021ecc}, i.e.,

\begin{equation}
\label{equ:7}
    T_{\rm merger}=\frac{5}{256} \frac{c^{5} a_{f}^{4}}{G^{3} (M_{\rm NS} + M_{\rm BH} )^{2} m_{r}} T(e) ,
\end{equation}
where $c$ is the speed of light and $m_r$ is the binary’s reduced mass, and the function $T(e)$ is given by
\begin{equation}
    T(e) = \left(1+0.27 e^{10}+0.33 e^{20}+0.2 e^{1000}\right)\left(1-e^{2}\right)^{7 / 2}.
\end{equation}

\bibliography{NSBH}{}
\bibliographystyle{aasjournal}

\end{document}